\documentclass[11pt,twoside]{article}
\usepackage{newpasp,epsf,psfig}
\def\edcomment#1{\iffalse\marginpar{\raggedright\sl#1\/}\else\relax\fi}
\reversemarginpar

\begin{document}
\title{The parsec-scale central components of FR I
radio galaxies}
\author{Preeti Kharb \& Prajval Shastri}
\affil{Indian Institute of Astrophysics, Koramangala,
Bangalore - 34, India.}

\begin{abstract}
A majority of a complete sample of 3CR FR I radio
galaxies show unresolved optical nuclear sources on
the scales of 0.$\arcsec$1. About half of the 3CR FR II 
radio galaxies observed with the HST also show Compact 
Central Cores (CCC). These CCCs have been interpreted as 
the optical counterparts of the non-thermal radio cores 
in these radio galaxies (Chiaberge, Capetti, \& Celotti 1999).
We show that the optical flux density of the CCCs in FR Is
is correlated with the radio core prominence. This 
correlation supports the argument of Chiaberge et al. 
that the CCC radiation is of a non-thermal
synchrotron origin, which is relativistically beamed
along with the radio emission.
\end{abstract}

\section{Introduction} 
{\it Radio Galaxies} are radio-loud $(\rm S(\nu)_{5 GHz}/S(\nu)_{B Band}>10)$
Active Galactic Nuclei (AGN) found in hosts that are 
elliptical galaxies. Their radio structure is made up of two lobes
of radio-emitting plasma situated on either side of an unresolved
radio core and connected to the core by plasma jets.
The radio morphologies fall in two distinct sub-classes:
the Fanaroff-Riley type I (FR I) with extended plumes and tails having
$L_{178}< 2\times 10^{26}$ W/Hz and the Fanaroff-Riley type II (FR II) 
with narrow jets and hotspots having $L_{178} > 2\times 10^{26}$ W/Hz 
at 178 MHz. 

Within the {\it unification scheme} for radio-loud AGN,
FR I and FR II radio galaxies are thought to represent the parent
populations of BL Lac objects and radio-loud quasars, respectively. 
BL Lac objects show clear evidence for relativistic beaming resulting from
the bulk relativistic motion of the plasma moving close to the line of sight.
If FR I radio galaxies are the plane-of-sky counterparts of BL Lacs,
FR I jets should also have bulk relativistic motion.

The {\it core prominence parameter} $R_{c}$, which is the ratio between
core and extended radio flux density, is a beaming indicator. This 
is because if the core is the unresolved relativistically beamed jet 
and the lobes are unbeamed then $R_{c}$ becomes a statistical measure 
of orientation. $R_{c}$ has indeed been shown to correlate
with other orientation-dependent properties in the case of FR IIs (eg.
Kapahi \& Saikia 1982) and in FR Is (eg. Laing et al. 1999).
The jet-to-counterjet surface brightness ratio $R_{j}$, is one
such parameter, where differences in the surface brightness
between the jet and counterjet at a given distance from the nucleus are
interpreted as effects of Doppler beaming and dimming respectively,
on intrinsically symmetrical flows.

The {\it primary motivation} for our work comes from the
discovery of unresolved optical nuclear components at the centres of
radio galaxies by the {\it Hubble Space Telescope}.
The optical flux of these Central Compact Cores (CCC),
(Chiaberge et al. 1999), show a striking linear correlation
with the radio core emission. Chiaberge et al. suggest that the CCCs are
the optical counterparts of the relativistic radio jet.
We study the relationship between the CCC flux densities of FR Is 
and FR IIs and the core prominence parameter, $R_{c}$, the 
jet-to-counterjet surface brightness ratios, $R_{j}$;
and the X-ray core flux densities.

\section{The sample and the data }
We used the sample of 27 3CR FR Is in Chiaberge et al. (1999),
with their data for the CCC flux densities. In addition, 
the CCC flux densities for the FR Is NGC 7052 and NGC 6251 are 
from Capetti \& Celotti (1999) and Hardcastle \& Worrall (2000) respectively.
The FR II radio galaxies and their CCC optical flux densities
are taken from the sample of 26 3CR FR IIs of Chiaberge et al.(2000).
The set of BL Lacs and their optical core flux densities
are from Capetti \& Celotti (1999).
The VLBI jet-to-counterjet surface brightness ratios
come from the VLBI observations of the sample of Giovannini et al. (1990).
(Giovannini et al. 1999 and references therein).
The jet-to-counterjet surface brightness ratio,
$R_j={B(jet)}/{B(cjet)}$,
are for distances of a few pc from the core.
The X-ray core flux densities are from the ROSAT data in 
Hardcastle \& Worrall (2000). The core prominence, 
$R_c ={S(core)}/{S(extended)}$,
was calculated using the total and core radio flux densities at 5 GHz.

\section{Results}
\subsection{Optical CCCs in FR I radio galaxies}
The CCC optical flux density $F_o$, is well correlated with radio 
core prominence $R_{c}$,
the significance of the correlation being at the 0.001 level
(Spearman rank test). See Figure 1a.
Given that $R_{c}$ is a statistical measure of angle to the line of sight,
the correlation suggests that the CCC flux density is orientation dependent
in the same way as the core radio emission and it may thus
constitute the optical counterpart of the radio synchrotron jet.
In Figure 1a we have included a set of BL Lacs, whose extended
radio luminosities span the same range as the FR Is, for comparison.
The BL Lacs extend the correlation to higher $R_{c}$, as would be expected
if FR Is are the parent population of BL Lacs.\\
The optical flux density of CCCs $F_o$, and the jet-to-counterjet 
surface brightness ratios $R_j$, at 5 GHz for FR I radio galaxies show a
correlation significant at the 0.01 level (Spearman rank correlation test).
See Figure 1b. This further substantiates the claim that the optical 
CCC is a part of the jet.

\subsection{Optical CCCs in FR II radio galaxies}
The optical core flux density for the sample of FR II radio galaxies 
shows a weaker correlation with radio core prominence than the 
FR Is (significance level 0.1; Spearman rank
correlation test). See Figure 2. 

\begin{figure}[h]
\plotfiddle{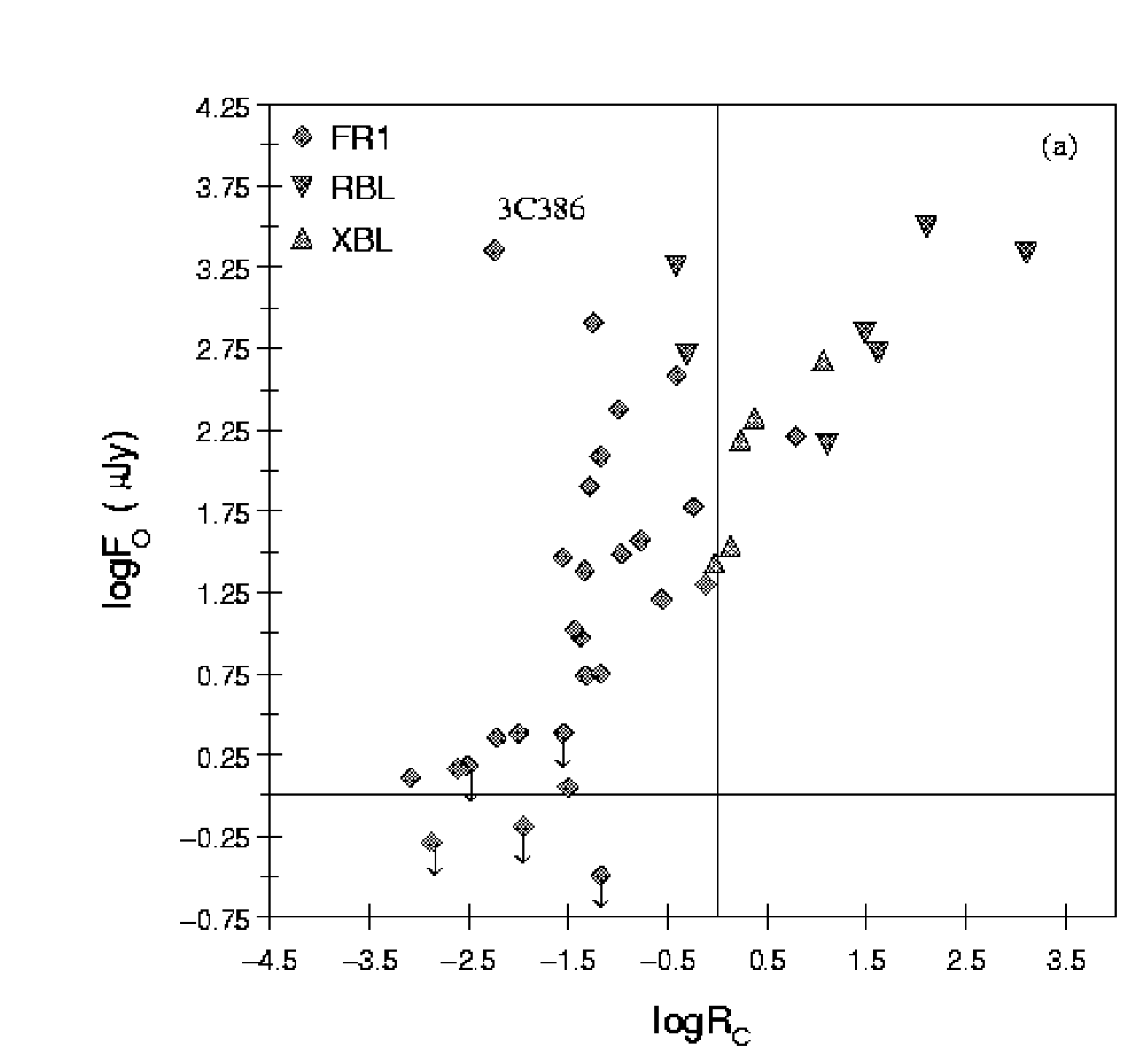}{2.77cm}{0}{35}{35}{-230}{-102}
\plotfiddle{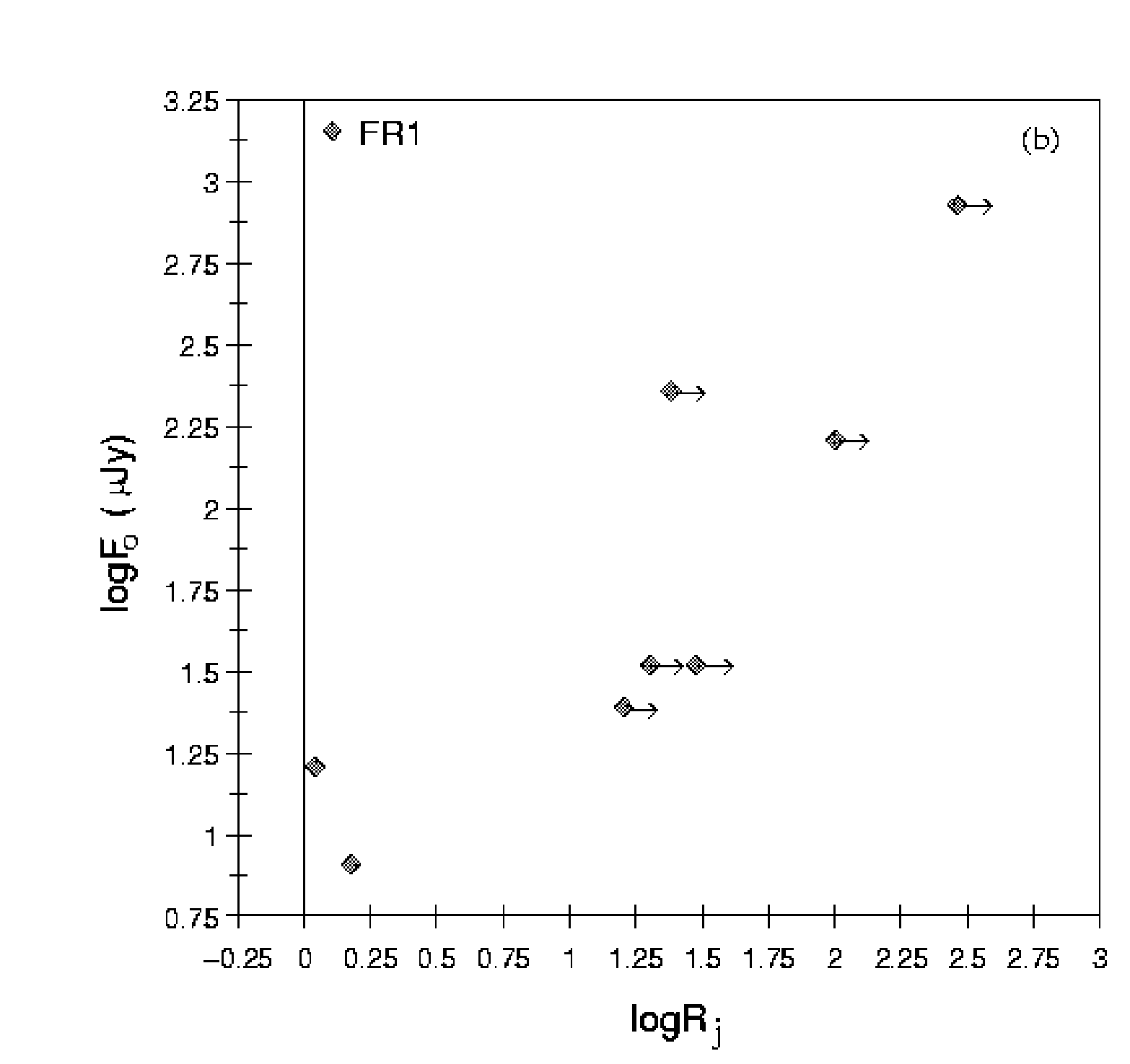}{2.77cm}{0}{35}{35}{-8}{-10}
\caption{(a). Radio core prominence, $R_{c}$ versus 
CCC flux density, $F_o$ $(\mu$Jy) for FR I radio galaxies and (RBL=radio-selected,
XBL=X-ray selected) BL Lacs. (b). Jet-to-counterjet ratio, $R_{j}$
versus $F_o$ $(\mu$Jy) for FR Is. (Arrows denote upper limits).}
\end{figure}

\subsection{X-ray core components in radio galaxies}
We examine the correlation between the X-ray 
core flux density and the radio core prominence. 
See Figure 3. This seems weaker than the previous correlations.
The correlation is significant at the 0.2 level for FR Is taken alone
while it is significant at the 0.1 level for FR I and FR IIs taken together.

\begin{figure}[h]
\plotfiddle{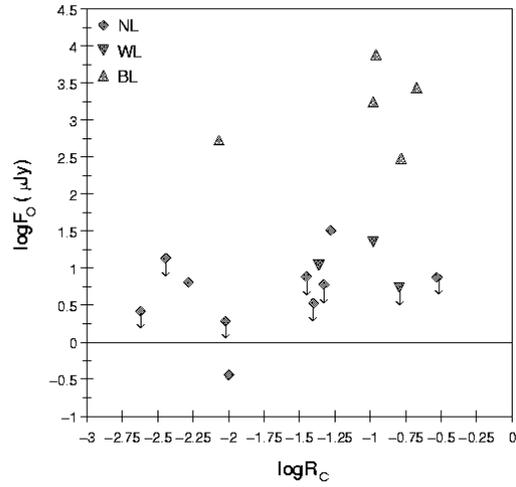}{5.22cm}{0}{35.5}{35.5}{-105}{-7}
\caption{Radio core prominence $R_c$, versus CCC 
flux density $F_o$ ($\mu$Jy), for 
FR II (NL=Narrow-line, WL=Weak-line, BL=Broad-line) radio galaxies. Arrows
denote upper limits in CCC flux density.}
\end{figure}

\begin{figure}[h]
\plotfiddle{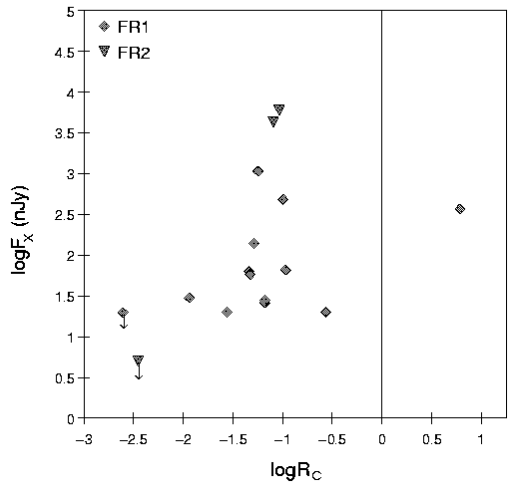}{5.22cm}{0}{35.5}{35.5}{-105}{-7}
\caption{Radio core-prominence $R_c$, versus X-ray core flux 
density $F_x$ (nJy), for FR I and FR II radio galaxies. Arrows 
denote upper limits in X-ray core flux density.}
\end{figure}

\end{document}